\documentclass[fleqn,reqno]{amsart}

\usepackage{graphicx,amscd,amsmath,amssymb,verbatim}
\usepackage[dvips]{hyperref}
\usepackage[TS1,OT1,T1]{fontenc}

\begin{document}

\title[Reciprocal relativistic quantum mechanics]{Reciprocally relativity of noninertial frames: quantum mechanics}
\author{Stephen G. Low}
\address{Austin, Texas }
\email{Stephen.Low@hp.com}
\date{\today}
\maketitle
\begin{abstract}

Noninertial transformations on time-position-momentum-energy space
$\{t,q,p,e\}$ with invariant Born-Green metric $d s^{2}=-d t^{2}+\frac{1}{c^{2}}
d q^{2}+\frac{1}{b^{2}}(d p^{2}-\frac{1}{c^{2}}d e^{2})$ and the
symplectic metric $-d e\wedge d t+ d p\wedge d q$ are studied. This
$\mathcal{U}1,3)$ group of transformations contains the Lorentz
group as the inertial special case and in the limit of small forces
and velocities, reduces to the expected Hamilton transformations
leaving invariant the symplectic metric and the nonrelativistic
line element $d s^{2}=-d t^{2}$. The $\mathcal{U}( 1,3) $ transformations
bound relative velocities by $c$ and relative forces by $b$.\ \ Spacetime
is no longer an invariant subspace but is relative to noninertial
observer frames. In the limit of $b\rightarrow \infty $, spacetime
is invariant. Born was lead to the metric by a concept of reciprocity
between position and momentum degrees of freedom and for this reason
we call this reciprocal relativity.\ \ \

For large $b$, such effects will almost certainly only manifest
in a quantum regime.\ \ Wigner showed that special relativistic
quantum mechanics follows from the projective representations of
the inhomogeneous Lorentz group. Projective representations of a
Lie group are equivalent to the unitary representations of its central
extension.\ \ The same method of projective representations for
the inhomogeneous $\mathcal{U}( 1,3) $ group is used to define the
quantum theory in the noninertial case. The central extension of
the inhomogeneous $\mathcal{U}( 1,3) $ group is the cover of the
quaplectic group $\mathcal{Q}( 1,3) =\mathcal{U}( 1,3) \otimes _{s}\mathcal{H}(4)$.\ \ $\mathcal{H}(
4) $ is the Weyl-Heisenberg group. The $\mathcal{H}( 4) $ group,
and the associated Heisenberg commutation relations central to quantum
mechanics, results directly from requiring projective representations.
A set of second order wave equations result from the representations
of the Casimir operators.\ \ 
\end{abstract}
\section{Introduction}
\subsection{Special relativity transformations}

Special relativity defines transformations between inertial frames
in spacetime.\ \ For simplicity of exposition, let us start by considering\ \ the
one dimensional case, $\{t,q\}\in \mathbb{M}\simeq \mathbb{R}^{2}$,
for which the global transforms on time and position are
\begin{equation}
\tilde{ t}=\gamma \mbox{}^{\circ}( v) \left( t+\frac{v}{c^{2}} q\right)
,\ \ \tilde{ q}=\gamma \mbox{}^{\circ}( v) \left(  q+v t\right)
.
\end{equation}

\noindent with
\begin{equation}
\gamma \mbox{}^{\circ}( v) ={\left( 1-\frac{v^{2}}{c^{2}}\right)
}^{-1/2}%
\label{gamma 0 definition}
\end{equation}

\noindent These transformations act locally on the frames $\{d t,
d q\}$ in the cotangent vector space
\begin{equation}
\begin{array}{l}
 d\tilde{ t}=\gamma \mbox{}^{\circ}( v) \left( d t+\frac{v}{c^{2}}d
q\right) , \\
 d\tilde{ q}=\gamma \mbox{}^{\circ}( v) \left(  d q+v d t\right)
.
\end{array}%
\label{basic special relativity local transformations}
\end{equation}

\noindent The corresponding transforms act on momentum-energy space
$\{p,e\}\in \tilde{\mathbb{M}}\simeq \mathbb{R}^{2}$\ \ 
\begin{equation}
\tilde{ p}=\gamma \mbox{}^{\circ}( v) \left( p+\frac{v}{c^{2}} e\right)
, \tilde{e}=\gamma \mbox{}^{\circ}( v) \left( e+v p\right) .
\end{equation}

\noindent In the neighborhood of an inertial frame, the local expressions
in terms of the frame $\{d p, d e\}$ in the cotangent vector space
is 
\begin{equation}
\begin{array}{l}
 d\tilde{ p}=\gamma \mbox{}^{\circ}( v) \left( d p+\frac{v}{c^{2}}
d e\right) , \\
 d\tilde{e}=\gamma \mbox{}^{\circ}( v) \left( d e+v d p\right) .
\end{array}%
\label{special relativity local momentum-energy transformations}
\end{equation}

\noindent These transformations leave invariant the orthogonal metrics
defining the line elements 
\begin{equation}
d s^{2} = -d t^{2}+\frac{1}{c^{2}}d q^{2} ,\ \ \ \ \ d \mu ^{2}=
d p^{2}-\frac{1}{c^{2}}d e^{2}.%
\label{metrics for special relativity}
\end{equation}

Spacetime $\mathbb{M}$ and momentum-energy space $\tilde{\mathbb{M}}$
may be combined to form the time-position-momentum-energy space\ \ $\mathbb{P}\simeq
\mathbb{M}\otimes \tilde{\mathbb{M}}\simeq \mathbb{R}^{4}$ and the
above expressions (3,5) may be regarded as acting on frames $d z=\{d
t, d q,d p, d e\} \in {T^{*}}_{z}\mathbb{P}$ in the cotangent vector
space of $\mathbb{P}$ where $z=\{t,q,p,e\}\in \mathbb{P}$.\ \ The
line elements\footnote{The component matrix of the line elements
are singular and therefore these line elements do not define metrics
on ${T^{*}}_{z}\mathbb{P}$.} given in (6) are defined on this space
that continue to be invariant under the action of (3,5).\ \ Additionally,
there is a symplectic metric $\zeta =-d e \wedge d t+d p\wedge d
q$ that is invariant under these transforms. 

The line elements $d s^{2}$ and ${\mathrm{d\mu }}^{2}$ are invariant
under\ \ $\mathcal{S}\mathcal{O}( 1,1) \otimes \mathcal{S}\mathcal{O}(
1,1) $\footnote{Each line element is actually invariant under $\mathcal{O}(
1,1) $. This group is required in the quantum case. }.\ \ The symplectic
metric $\zeta$ is invariant under\ \ the symplectic group $\mathcal{S}p(
4) $.\ \ \ Transformations leaving both the line elements and the
symplectic metric invariant are in the intersection of these of
these two groups 
\begin{equation}
\left( \mathcal{S}\mathcal{O}( 1,1) \otimes \mathcal{S}\mathcal{O}(
1,1) \right)  \cap \mathcal{S}p( 4)  \simeq \mathcal{S}\mathcal{O}(
1,1) .%
\label{special relativity group intersection 1d}
\end{equation}

This $\mathcal{S}\mathcal{O}( 1,1) $ group is the group of local
inertial canonical\footnote{Transformations leaving the symplectic
metric invariant are generally referred to as {\itshape canonical}.
} transformations on ${T^{*}}_{z}\mathbb{P}$.\ \ Elements in the
group may be written as the $4\times 4$ real matrix group $\Lambda
( v) $ with\ \ transformations
\begin{equation}
d\tilde{z}=\frac{\partial \tilde{z}}{\partial z } d z= \Lambda (
v) d z,%
\label{special relativity lambda transform}
\end{equation}

\noindent (using matrix notation,) where the matrices are explicitly
\begin{equation}
\Lambda ( v) =\gamma \mbox{}^{\circ}( v) \left( \begin{array}{llll}
 1  & \frac{v}{c^{2}} & 0 & 0 \\
  v  & 1  & 0 & 0 \\
 0 & 0 & 1  & \frac{v}{c^{2}} \\
 0 & 0 & v  & 1 
\end{array}\right) .%
\label{special relativity lambda matrix}
\end{equation}

It is the symplectic condition in (7) that requires that the velocity
parameter for the independent\ \ $\mathcal{S}\mathcal{O}( 1,1) $
groups in the direct product to be the same. The transformations
by $\Lambda ( v) $ in (8)\ \ are the transformations given in (3,5).
The group multiplication law is the usual relation for the addition
of velocities in this one dimensional case of special relativity\ \ 
\begin{equation}
\Lambda ( v) \cdot \Lambda ( \tilde{v}) =\Lambda ( \frac{v+\tilde{v}}{1+v\tilde{v}/c^{2}})
.%
\label{special relativity velocity addition}
\end{equation}

Time is relative to the inertial observer in special relativity.
There is no absolute rest frame. However, implicit in the restriction
that these transformations are valid only between inertial observers
is the assumption of an absolute inertial frame that all observers
agree on.\ \ The velocity addition law ensures that the addition
of velocities is bounded by $c$.\ \ 

The above discussion generalizes straightforwardly to the $1+3$
dimensional case for which the group defined in (7) becomes
\begin{equation}
\left( \mathcal{S}\mathcal{O}( 1,3) \otimes \mathcal{S}\mathcal{O}(
1,3) \right)  \cap \mathcal{S}p( 8)  \simeq \mathcal{S}\mathcal{O}(
1,3) .
\end{equation}

\noindent All of the groups in this expression are matrix groups\ \ and
therefore the group elements are conveniently realized as $8$ dimensional
matrices.
\subsection{Nonrelativistic inertial transformations: Hamilton's
equations}

The nonrelativistic limit is the case where $v/c \rightarrow 0$
or equivalently $c\rightarrow \infty $.\ \ In this limit, (3) and\ \ (5)
reduce to
\begin{equation}
\begin{array}{l}
 d\tilde{t}=d t, \\
 d\tilde{q}=d q + v d t, \\
 d\tilde{p}=d p, \\
 d \tilde{e} = d e + v d p,
\end{array}%
\label{Hamilton inertial transforms}
\end{equation}

\noindent and in the limit $c\rightarrow \infty $ the\ \ line elements
in (6) reduce to
\begin{equation}
d s^{2} = -d t^{2} ,\ \ \ \ \ d \mu ^{2}= d p^{2}.%
\label{nonrelativistic line elements}
\end{equation}

The symplectic metric $\zeta $ is not affected by the limit and
continues to be an invariant of the transformations.\ \ The group
leaving both these invariant is the contraction of the one dimensional
Lorentz group to the Euclidean group.\ \ 
\begin{equation}
\operatorname*{\lim }\limits_{c\,\rightarrow \:\infty }\mathcal{S}\mathcal{O}(
1,1) =\mathcal{E}( 1) \simeq \mathcal{T}( 1) .
\end{equation}

\noindent As the one dimensional Euclidean group is the translation
group, the contraction of the matrix realization in\ \ (7)\ \ is
\begin{equation}
\Phi ( v) =\operatorname*{\lim }\limits_{c\,\rightarrow \:\infty
}\Gamma ( v)  =\left( \begin{array}{llll}
 1  & 0 & 0 & 0 \\
  v  & 1  & 0 & 0 \\
 0 & 0 & 1  & 0 \\
 0 & 0 & v  & 1 
\end{array}\right) .%
\label{Hamilton matrix element phi}
\end{equation}

\noindent This satisfies the group composition law for translation
group, $\Phi ( \tilde{v}) \cdot \Phi ( v)  =\Phi ( \tilde{v}+v)
$.\ \ 

The transformations in (12) are the canonical transformations between
inertial frames in non-relativistic Hamilton mechanics.\ \ Because
time in is an invariant of these transformations, we say that time
is invariant or {\itshape absolute; }all inertial observers agree
on the definition of the time subspace of this time-position-momentum-energy
space. Velocity is simply additive and is unbounded. It follows
from the transformations that there is a special frame that is an
absolute inertial rest frame.

The canonical transformations in the nonrelativistic limit may be
integrated to determine the transformations\ \ ${\tilde{z}}^{a}=f^{a}(
z) $,\ \ 
\begin{equation}
d{\tilde{z}}^{a}=\frac{\partial f^{a}( z) }{\partial z^{b} } d z^{b}=
{\left[ \Phi ( v) \right] }_{b}^{a}d z^{b},\ \ \ \ \ \ \ \frac{\partial
f^{a}( z) }{\partial z^{b} }= {\left[ \Phi ( v) \right] }_{b}^{a}.%
\label{jacobian of tx}
\end{equation}

To explicitly compute these partials of $f^{a}( t,q,p,e) $ with
$z=\{t,q,p,e\}$, use the form of $\Phi ( v) $ given in (22).\ \ The
diagonal elements are the boundary conditions
\begin{equation}
\frac{\partial f^{1}}{\partial t }=1,\ \ \frac{\partial f^{2}}{\partial
q }=1,\ \ \frac{\partial f^{3}}{\partial p }=1,\ \ \frac{\partial
f^{4}}{\partial e }=1,
\end{equation}

\noindent and the remaining nonzero terms are Hamilton's equation
for the velocity
\begin{equation}
\frac{\partial f^{2}}{\partial t }=v=\frac{\partial f^{4}}{\partial
p }.%
\label{Hamilton inertial velocity equation}
\end{equation}

\noindent All other partials are zero, including what would be the
second of Hamilton's equations for forces, as one would expect for
an inertial transformation 
\begin{equation}
\frac{\partial f^{3}}{\partial t }=f=0=-\frac{\partial f^{4}}{\partial
q },\ \ \ \ \frac{\partial f^{4}}{\partial t }=r=0.%
\label{Hamilton inertial force equation}
\end{equation}

\noindent These nonrelativistic equations may be integrated, neglecting
trivial constants, to define the inertial canonical transformations$\text{}$
${\tilde{z}}^{a}=f^{a}( z) $.\ \ \ 
\begin{equation}
\begin{array}{l}
 \tilde{t}=f^{1}( t,q,p,t) =t, \\
 \tilde{q}=f^{2}( t,q,p,t) =q+q( t) =q+v t, \\
 \tilde{p}=f^{3}( t,q,p,t) =p, \\
 \tilde{e}=f^{4}( t,q,p,t) =e+H( p) .
\end{array}%
\label{Hamilton global transformations}
\end{equation}

\noindent Then, (18,19) are 
\begin{equation}
\frac{d q( t) }{d t }=v=\frac{\partial H}{\partial p }, \frac{d
p( t) }{d t }=f=-\frac{\partial H}{\partial q },\ \ \ \ \frac{\partial
H}{\partial t }=r.%
\label{Hamilton equations}
\end{equation}

\noindent with $f=r=0$ for this inertial case. 
\subsection{Nonrelativistic noninertial transformations: Hamilton's
equations}

Hamilton's equations and the corresponding canonical transformations
are generally valid for noninertial transformations where forces
are non-zero.\ \ A frame associated with an arbitrary particle obeying
Hamilton's equations is generally noninertial.\ \ In this case,
the equations in\ \ (19)\ \ are no longer zero and consequently
the matrix (15) becomes
\begin{equation}
\Phi ( v,f,r)  =\left( \begin{array}{llll}
 1  & 0 & 0 & 0 \\
  v  & 1  & 0 & 0 \\
 f & 0 & 1  & 0 \\
 r & -f & v  & 1 
\end{array}\right) .%
\label{Hamilton general matrix phi}
\end{equation}

By direct matrix multiplication, it may be verified that this is
a matrix group with product \footnote{This is the group composition
law for the Weyl-Heisenberg group. The reason for this is discussed
following (45)} 
\begin{equation}
\Phi ( \tilde{v},\tilde{f},\tilde{r})  \cdot \Phi ( v,f,r)  =\Phi
( v+\tilde{v},f+\tilde{f},r+\tilde{r}+v\tilde{f}-f\tilde{v}) .%
\label{Hamilton group composition law}
\end{equation}

\noindent We call this group the Hamilton group $\mathcal{H}a( 1)
$. This leads to the set of transformations $d\tilde{z}= \Phi (
v,f,r) d z$ that are explicitly
\begin{equation}
\begin{array}{l}
 d\tilde{t}=d t, \\
 d\tilde{q}=d q + v d t, \\
 d\tilde{p}=d p+f d t, \\
 d \tilde{e} = d e + v d p-f d q +r d t.
\end{array}%
\label{Hamilton general transformation equaitons}
\end{equation}

These leave invariant the line element $d s^{2}=-d t^{2}$ and the
symplectic metric $\zeta =-d e \wedge d t+d p\wedge d q$.\ \ As
expected, the line element ${\mathrm{d\mu }}^{2}= d p^{2}$ is no
longer an invariant.\ \ \ Again, these may be integrated to give
\begin{equation}
\begin{array}{l}
 \tilde{t}=f^{1}( t,q,p,t) =t, \\
 \tilde{q}=f^{2}( t,q,p,t) =q+q( t) , \\
 \tilde{p}=f^{3}( t,q,p,t) =p+p( t) , \\
 \tilde{e}=f^{4}( t,q,p,t) =e+H( p,q,t) ,
\end{array}%
\label{Hamilton noinertial transformations}
\end{equation}

\noindent Using (16)\ \ this directly results in Hamilton's equations
that are given in (21) with $f,r$ not necessarily zero

Alternatively, one can start from the assumption that the line element\ \ $d
s^{2}=-d t^{2}$ and the symplectic metric are invariant and arrive
at the matrix group (22).\ \ This, in turn, leads directly to the
transformation equations (24,25)\ \ and Hamilton's equations\ \ (21).\ \ 

This establishes the equivalence of the formulations.\ \ These equations
and arguments readily generalize to $3$ spacial dimensions.

The transformations in (20, 24) are the canonical transformations
between\ \ frames that are generally noninertial for non-relativistic
Hamilton mechanics. As in the inertial case, time in is an invariant
of these transformations and so we say that time is invariant or
{\itshape absolute.\ \ \ }All observers agree on the definition
of the time subspace of this time-position-momentum-energy space.
Velocity is simply additive and is unbounded.\ \ Force is simply
additive and unbounded.\ \ It follows from the transformations that
there is a special frame that is an absolute inertial rest frame.
\subsection{Noninertial relativistic transformations }

The Lorentz group\ \ is the group of transformations between inertial
frames in special relativity.\ \ \ Hamilton's group is the group
of noninertial transformations in nonrelativistic Hamilton's mechanics.\ \ Both
of these groups leave invariant the symplectic metric $\zeta $ and
are therefore are subgroups of the symplectic group.\ \ Again, special
relativity leaves invariant the line elements 
\begin{equation}
d s^{2} = -d t^{2}+\frac{1}{c^{2}}d q^{2} ,\ \ \ \ \ d \mu ^{2}=
d p^{2}-\frac{1}{c^{2}}d e^{2}%
\label{special relativity line element}
\end{equation}

\noindent whereas Hamilton's equations leaves invariant the nonrelativistic
line element $d s^{2}=-d t^{2}$.\ \ We are looking for the group
of transformations that reduces to the special relativistic transformations
in the special case of inertial frames.\ \ Furthermore in the limit
of small velocities and small forces, it must contract to Hamilton's
group (22).\ \ A group with this property follows directly by combining
the degenerate orthogonal line elements into the single orthogonal
Born-Green metric \cite{born2-1} 
\begin{equation}
d s^{2} = -d t^{2}+\frac{1}{c^{2}}d q^{2} +\ \ \frac{1}{b^{2}}\left(
d p^{2}-\frac{1}{c^{2}}d e^{2}\right) %
\label{Born-Green metric}
\end{equation}

\noindent $b$ is a universal constant, and together with $c$, and
$\hbar $, defines a dimensional basis\footnote{$G$ is usually used
as the third dimensional constant, $G=\alpha _{G}\frac{c^{4}}{b}$.\ \ If
$\alpha _{G}$ turns out to be unity, then these are the usual Planck
scales.\ \ } with scales
\begin{equation}
\lambda _{t}=\sqrt{\frac{\hbar }{b c}}, \lambda _{q}=\sqrt{\frac{\hbar
c}{b}}, \lambda _{p}=\sqrt{\frac{\hbar  b}{ c}}, \lambda _{e}=\sqrt{\hbar
b c}.
\end{equation}

\noindent The group that leaves this orthogonal metric invariant
is $\mathcal{O}( 2,2) $. The symplectic metric $\zeta $ continues
to be invariant.\ \ \ \ The group of transformations leaving both
the orthogonal and symplectic metric invariant is 
\begin{equation}
\mathcal{O}( 2,2)  \cap \mathcal{S}p( 4)  \simeq \mathcal{U}( 1,1)
.%
\label{unitary group intersection}
\end{equation}

\noindent The unitary group is basic to the quantum formulation.
Again, as in (7), in the classical case, only the special orthogonal
metric needs to be considered and therefore
\begin{equation}
\mathcal{S}\mathcal{O}( 2,2)  \cap \mathcal{S}p( 4)  \simeq \mathcal{S}\mathcal{U}(
1,1) .%
\label{Classical case of group intersection}
\end{equation}

The group elements of $\mathcal{S}\mathcal{U}( 1,1) $\ \ may be
realized as the matrices $\Gamma ( v,f,r) $
\begin{equation}
\Gamma ( v,f,r) =\gamma ( v,f,r)  \left( \begin{array}{llll}
 1  & \frac{v}{ c^{2}} & \frac{f}{ b^{2}} & -\frac{r}{b^{2} c^{2}}
\\
 v & 1 & \frac{r}{b^{2}} & \frac{-f}{ b^{2}} \\
 f & -\frac{r}{c^{2} } & 1 & \frac{v}{c^{2}\ \ } \\
 r & -f &  v  & 1 
\end{array}\right) ,%
\label{quaplectic group v f r matrix}
\end{equation}

\noindent with $\gamma ( v,f,r) ={(1-v^{2}/c^{2}- f^{2}/b^{2}\mathrm{+}r^{2}/b^{2}
c^{2})}^{-1/2}$.\ \ 
\subsubsection{Reciprocal relativity transformation equations}

This group defines the transformations\ \ $d\tilde{z}= \Gamma (
v,f,r) d z$ that are explicitly
\begin{equation}
\begin{array}{l}
 d\tilde{ t}=\gamma ( d t+\frac{v}{c^{2}}d q+\frac{f}{b^{2}}d p-
\frac{r}{b^{2}c^{2}}d e) , \\
 d\tilde{ q}=\gamma (  d q+v d t+\frac{r}{b^{2}}d p-\frac{f}{b^{2}}d
e) . \\
 d\tilde{ p}=\gamma ( d p+ f d t-\frac{r}{c^{2}} d q +\frac{v}{c^{2}}
d e) , \\
 d\tilde{e}=\gamma ( d e+v d p-f d q+r d t) .
\end{array}%
\label{u13 transformation equations}
\end{equation}

\noindent The group composition law between three frames is 
\begin{equation}
\Gamma ( v^{{\prime\prime}},f^{{\prime\prime}},r^{{\prime\prime}})
\cdot \Gamma ( v^{\prime } ,f^{\prime },r^{\prime }) =\Gamma ( v,f,r)
\label{quaplectic group composition}
\end{equation}

\noindent with $v=g_{v}( v^{\prime },v^{{\prime\prime}},f^{\prime
},f^{{\prime\prime}},r^{\prime },r^{{\prime\prime}}) $,\ \ $f=g_{f}(
v^{\prime },v^{{\prime\prime}},f^{\prime },f^{{\prime\prime}},r^{\prime
},r^{{\prime\prime}}) $ and $r=g_{r}( v^{\prime },v^{{\prime\prime}},f^{\prime
},f^{{\prime\prime}},r^{\prime },r^{{\prime\prime}}) $ where these
are given by
\begin{equation}
\begin{array}{l}
 v= \left( v^{{\prime\prime}}+v^{\prime }\mathrm{+}\frac{1}{b^{2}}\left(
r^{\prime }f^{{\prime\prime}}-f^{\prime }r^{{\prime\prime}}\right)
\right) /\left( 1+ \frac{v^{\prime }v^{{\prime\prime}}}{c^{2}}+
\frac{f^{\prime }f^{{\prime\prime}}}{b^{2}}\mathrm{-}\frac{r^{\prime
} r^{{\prime\prime}}}{b^{2} c^{2}}\right) , \\
 f= \left( f^{{\prime\prime}}+f^{\prime }\mathrm{+}\frac{1}{c^{2}}\left(
-r^{\prime }v^{{\prime\prime}}+v^{\prime }r^{{\prime\prime}}\right)
\right) /\left( 1+ \frac{v^{\prime }v^{{\prime\prime}}}{c^{2}}+
\frac{f^{\prime }f^{{\prime\prime}}}{b^{2}}\mathrm{-}\frac{r^{\prime
} r^{{\prime\prime}}}{b^{2} c^{2}}\right) , \\
 r=\left( r^{{\prime\prime}}+r^{\prime }-f^{\prime }v^{{\prime\prime}}+v^{\prime
}f^{{\prime\prime}}\right) /\left( 1+ \frac{v^{\prime }v^{{\prime\prime}}}{c^{2}}+
\frac{f^{\prime }f^{{\prime\prime}}}{b^{2}}\mathrm{-}\frac{r^{\prime
} r^{{\prime\prime}}}{b^{2} c^{2}}\right) 
\end{array}%
\label{quaplectic relativity rate transformations}
\end{equation}

Proper acceleration at a specific moment is defined relative to
a frame that may be taken to be $(v^{{\prime\prime}},f^{{\prime\prime}},r^{{\prime\prime}})$
with $(d v^{{\prime\prime}},d f^{{\prime\prime}},d r^{{\prime\prime}})=0$
\cite{Rindler}.\ \ The primed frame is momentarily locally inertially
at rest relative to this frame, $(v^{\prime },f^{\prime },r^{\prime
})=0$,\ \ but has non-zero derivatives $(d v^{\prime },d f^{\prime
},d r^{\prime })$. Therefore, from (34), at this moment,\ \ \ $(\mathit{v}\mathit{,}\mathit{f}
,\mathit{r} )=(v^{{\prime\prime}},f^{{\prime\prime}},r^{{\prime\prime}})$.\ \ Then,
taking the derivative of (33) with these conditions gives\ \ \ 
\begin{equation}
\begin{array}{l}
 d v=\left( 1-\frac{v^{2}}{c^{2}}\right) d v^{\prime }+\frac{1}{b^{2}}\left(
\left( f+\frac{r v}{ c^{2}}\right)  d r^{\prime } - \left( r+f v\right)
{df}^{\prime }\right) , \\
 d f= \left( 1-\frac{f^{2}}{b^{2}}\right)  {\mathrm{df}}^{\prime
}+\frac{1}{c^{2}}\left( \left( -v+\frac{f r}{b^{2} }\right)  d r^{\prime
} + \left( r-f v\right) d v^{\prime }\right) , \\
 d r=\left( 1+\frac{r^{2}}{b^{2} c^{2}}\right)  d r^{\prime } -\left(
v+\frac{f r}{b^{2}}\right)  d f^{\prime }+\left( f-\frac{v r}{c^{2}}\right)
d v^{\prime }.
\end{array}%
\label{quaplectic relativity rate differential}
\end{equation}

\noindent Noting that $d t^{\prime }= {\gamma ( v,f,r)  }^{-1}d
t$, the derivative with respect to $d t$ may be computed and the
equations inverted to yield the transformation of proper acceleration
and impulse
\begin{equation}
\begin{array}{l}
 \frac{d v^{\prime }}{d t^{\prime }}=\gamma {\left( v,f,r\right)
}^{3} \left(  \frac{d \mathit{v}}{d\mathit{t}}+\frac{1}{b^{2}}\left(
f \frac{d \mathit{r}}{d \mathit{t}}-r \frac{d \mathit{f}}{d \mathit{t}}
\right) \right) , \\
 \frac{d f^{\prime }}{d t^{\prime }}=\gamma {\left( v,f,r\right)
}^{3} \left(  \left( \frac{d \mathit{f}}{d \mathit{t}}+\frac{1}{c^{2}}\left(
v\ \ \frac{d \mathit{r}}{d \mathit{t}}-r \frac{d \mathit{v}}{d\mathit{t}}\right)
\right) ,\right.  \\
 \frac{d r^{\prime }}{d t^{\prime }}=\gamma {\left( v,f,r\right)
}^{3} \left(  \frac{d \mathit{r}}{d \mathit{t}}- f\frac{d \mathit{v}}{d\mathit{t}}+v
\frac{d \mathit{f}}{d \mathit{t}}\right) .
\end{array}%
\label{quaplectic acceleration equations}
\end{equation}

These transformations have the\ \ property that for inertial transformations
where $f=r=0$, that $\Gamma ( v,0,0) =\Lambda ( v) $ where $\Lambda
( v) $ is the special relativity transformation in (9). The special
case $f=r=0$ of the velocity transformation (34) and proper acceleration
(36) are \ \ 
\begin{equation}
\begin{array}{l}
 v=g \mbox{}^{\circ}_{v}( v^{\prime } ,v^{{\prime\prime}}) =\left(
v^{\prime }+v^{{\prime\prime}}\right) /\left( 1+ \frac{v^{\prime
} v^{{\prime\prime}}}{c^{2}}\right) , \\
 \frac{d v^{\prime }}{d t^{\prime }}= {\gamma \mbox{}^{\circ}( v)
}^{3} \frac{d \mathit{v}}{d \mathit{t}},
\end{array}%
\label{special relativity velocity acceleration transformation}
\end{equation}

\noindent where $\gamma ( v,0,0) =\gamma \mbox{}^{\circ}( v) $ is
defined in (2). In this case, the velocity transformation is identical
to the usual special relativity expression (10) and proper acceleration
expression in (37) are as expected in \cite{Rindler}

Null surfaces separate timelike from spacelike trajectories.\ \ In
one dimensional special relativity, these are simply the cones $\frac{1}{c^{2}}d
q^{2}=d t^{2}$\ \ or $v=\pm c$.\ \ It follows directly that the
velocity addition law has the fixed point $v=g \mbox{}^{\circ}_{v}(
v,v) |_{v=\pm c}$.\ \ A fixed point surface for the noninertial
transformations (34)\ \ that have the property that $v=g_{v}( \mathit{v}
,v,f,f,r,r) $, $f=g_{f}( \mathit{v} ,v,f,f,r,r) $ and $r=g_{r}(
\mathit{v} ,v,f,f,r,r) $ is\footnote{Additional branches of the
null surface exist for which $r\neq 0$ that require further investigation}
\begin{equation}
\frac{v^{2}}{c^{2}}+\frac{f^{2}}{b^{2}}=1,\ \ \ \ r=0.\text{}%
\label{quaplectic null surface}
\end{equation}

\noindent In this case, the four dimensional $(q,p,e,t)$ space may
be visualized as three dimensional slices $(q,p,t)$ with $e $ constant.\ \ The
null surfaces are the cones $\frac{1}{c^{2}}d q^{2}+\frac{1}{b^{2}}d
p^{2}=d t^{2}$. In the inertial case with $d p=0$, these reduce
to the special relativity case $v=\pm c$.\ \ There is\ \ the corresponding
case where the velocity is zero where $f=\pm b$. 

Time is clearly not an invariant subspace of the transformations
and therefore time is relative to the observer frame as is the case
in special relativity.\ \ In addition, these transformations do
not have position-time (or {\itshape spacetime}) as an invariant
subspace of the group of transformations. This means that spacetime
is relative to the frames of noninertial observers. These effects
become significant for relative forces between particle states that
are large and approach the limiting value $b$ corresponding to the
extreme noninertial case. Thus, we have the phenomena that the transformations
mix the time-position with the energy-momentum degrees of freedom.\ \ Spacetime
itself has become relative.\ \ 
\subsubsection{Special relativity and nonrelativistic limits}

The special relativistic limit is the case where forces are small
relative to the scale $b$,\ \ $f/b\rightarrow 0$ . This is equivalent
to the\ \ limit, $b\rightarrow \infty $. In this limit, the Born-Green
line element defined in (27) reduces to the relativistic line element
in (26), $ -d t^{2}+\frac{1}{c^{2}}d q^{2} $.\ \ Furthermore, the
transformation equations (32) with the corresponding velocity and
proper acceleration equations given in (37)\ \ 
\begin{equation}
\begin{array}{l}
 d\tilde{ t}=\gamma \mbox{}^{\circ}( d t+\frac{v}{c^{2}}d q) , \\
 d\tilde{ q}=\gamma \mbox{}^{\circ}(  d q+v d t) . \\
 d\tilde{ p}=\gamma \mbox{}^{\circ}( d p+ f d t+\frac{1}{c^{2}}\left(
v d e-r d q \right) ) , \\
 d\tilde{e}=\gamma \mbox{}^{\circ}( d e+v d p-f d q+r d t) .
\end{array}%
\label{quapletic tx b inf limit}
\end{equation}

\noindent This defines the matrix group
\begin{equation}
\operatorname*{\lim }\limits_{b\,\rightarrow \:\infty } \Gamma (
v,f,r) =\Gamma \mbox{}^{\circ}( v,f,r) = \gamma \mbox{}^{\circ}(
v)  \left( \begin{array}{llll}
 1  & \frac{v}{ c^{2}} & 0 & 0 \\
 v & 1 & 0 & 0 \\
 f & -\frac{r}{c^{2} } & 1 & \frac{v}{ c^{2}} \\
 r & -f &  v  & 1 
\end{array}\right) ,%
\label{quaplectic group b limit v f r matrix}
\end{equation}

\noindent The corresponding velocity transformation that is obtained
from the group composition law $\Gamma \mbox{}^{\circ}( v^{{\prime\prime}},f^{{\prime\prime}},r^{{\prime\prime}})
\cdot \Gamma \mbox{}^{\circ}( v^{\prime } ,f^{\prime },r^{\prime
}) =\Gamma \mbox{}^{\circ}( v,f,r) $ and proper acceleration equations
are of the form given in (37). In addition, the force and power
transformations are the expected
\begin{equation}
\begin{array}{l}
 f=g \mbox{}^{\circ}_{f}( v^{\prime },v^{{\prime\prime}},f^{\prime
},f^{{\prime\prime}},r^{\prime },r^{{\prime\prime}}) =\left( f^{{\prime\prime}}+f^{\prime
}\mathrm{+}\frac{1}{c^{2}}\left( v^{\prime } r^{{\prime\prime}}-r^{\prime
}v^{{\prime\prime}}\right)  \right) /\left( 1+\frac{v^{\prime }v^{{\prime\prime}}}{c^{2}}\right)
, \\
 r=g \mbox{}^{\circ}_{r}( v^{\prime },v^{{\prime\prime}},f^{\prime
},f^{{\prime\prime}},r^{\prime },r^{{\prime\prime}}) =\left(  r^{{\prime\prime}}+r^{\prime
}\mathit{-}f^{\prime }v^{{\prime\prime}}+v^{\prime }f^{{\prime\prime}}
\right) /\left( 1+\frac{v^{\prime }v^{{\prime\prime}}}{c^{2}}\right)
.
\end{array}%
\label{quaplectic relativity b limit rate transformations}
\end{equation}

The nonrelativistic limit is now both small velocities, $v/c$$ \rightarrow
$0, and small forces, $f/b\rightarrow 0$. This is equivalent to
the limit $b,c\rightarrow \infty $.\ \ In this limit, the Born-Green
line element $d s^{2}$ defined in (27) reduces to the nonrelativistic
line element (13).\ \ Furthermore, 
\begin{equation}
\operatorname*{\lim }\limits_{b,c\,\rightarrow \:\infty }\Gamma
( v,f,r) = \Phi ( v,f,r) ,
\end{equation}

\noindent and therefore the transformations reduce in this limit
to\ \ Hamilton's equations and the associated transformations (24).

 These equations may readily be generalized to the $1+3$ dimensional
case in which case the group is\ \ \cite{Low3} 
\begin{equation}
\mathcal{S}\mathcal{O}( 2,6)  \cap \mathcal{S}p( 8)  \simeq \mathcal{S}\mathcal{U}(
1,3) .
\end{equation}

\noindent The special relativity\ \ limiting form is 
\begin{equation}
\operatorname*{\lim }\limits_{b\,\rightarrow \:\infty }\mathcal{S}\mathcal{U}(
1,3) =\mathcal{S}\mathcal{O}( 1,3) \otimes _{s}\mathcal{A}b( 4)
\end{equation}

\noindent where $\mathcal{A}b( 4) $ is a $4(4+1)/2=10$ dimensional
abelian group whose generators transform under the action of the
Lorentz generators as a (0,2) symmetric tensor\ \ and\ \ physically
correspond to\ \ ``force-power stress''.

Again, in the special relativistic limit with $b\rightarrow \infty
$, the position-time degrees of freedom no longer mix with the energy-momentum
degrees of freedom and an absolute position-time, or {\itshape spacetime,
} subspace that all\ \ observers agree on is recovered. This is
analogous to the recovery of an absolute concept of time in the
$c\rightarrow \infty $ limit of special relativity. 

The nonrelativistic limiting form is 
\begin{equation}
\text{}\operatorname*{\lim }\limits_{c,b\,\rightarrow \:\infty }\mathcal{S}\mathcal{U}(
1,3) =\mathcal{H}a( 3) = \mathcal{S}\mathcal{O}( 3) \otimes _{s}\mathcal{H}(
3) 
\end{equation}

\noindent where $\mathcal{H}( 3) $ is the Weyl-Heisenberg group.
(Equation (17) in \cite{Low3}). Note that the corresponding\ \ limit\ \ $b,c\rightarrow
\infty $ of $\mathcal{S}\mathcal{U}( 1,1) $ is $\mathcal{H}( 1)
$ and therefore $\mathcal{H}a( 1) \simeq \mathcal{H}( 1) $ as given
in (22, 23).

$b$ is defined in terms of $G$ as $b=\alpha _{G}\frac{c^{4}}{G}\approx
({10}^{44 }N)\ \ \alpha _{G}$ . If $\alpha _{G}$ is within a few
orders of unity, then the forces at which this occurs are very large.\ \ Such
forces between particle states would exist is the very early universe
where interactions are very strong and frames are strongly noninertial.

Born \cite{born2-1,born1}\cite{born2-1}was led to the Born-Green
metric through a principle of reciprocity that sought to make the
form of the physical equations invariant under the transform $\{q,p\}\rightarrow
\{p,-q\}$ and $\{t,e\}\rightarrow \{-e,t\}$ \cite{Low5}.\ \ It can
be verified that these transforms are a discrete automorphism of
this group. For this reason, we call the relativity of noninertial
frames described above, {\itshape reciprocal relativity.\ \ }
\section{Relativistic quantum mechanics}
\subsection{Special relativistic quantum mechanics.}

Particle states in quantum theory are represented by rays $\Psi$
in a Hilbert space $\text{\boldmath $\mathrm{H}$}$.\ \ Rays are
equivalence classes of states $|\psi \rangle \in \text{\boldmath
$\mathrm{H}$}$ defined up to a phase, $|\tilde{\psi }\rangle \simeq
|\psi \rangle $ if $|\tilde{\psi }\rangle =e^{i \omega }|\psi \rangle
$ with $\omega \in \mathbb{R}$. Rays are transformed from one to
another through projective transformations that are unitary (or
antiunitary)\ \ transformations up to a phase. 

Due to Wigner's work \cite{wigner},\ \ special relativistic quantum
mechanics is now understood in terms of the projective representations
of the inhomogeneous Lorentz group.\ \ Projective representations
are equivalent to the unitary, (or antiunitary\footnote{Antiunitary
representations are required for the extended group that includes
the discrete automorphisms. We refer here-on only to unitary representations
and leave this understood. }), representations $\varrho $ of the
central extension of this group \cite{Weinberg1}.\ \ \ As described
in Appendix A, central extensions arise either algebraically through
the addition of essential generators to the center of the algebra
that conform to the Jacobi identities, or topologically where the
group is lifted to its universal cover and the central elements
are the first homotopy group. Mackey's method for semidirect product
groups may be used to determine the unitary irreducible representations
(see Appendix B) \cite{mackey,Major,Low}. 

The special relativity line element (7) is invariant under $\mathcal{O}(
1,3) \simeq \mathcal{D}_{4}\otimes _{s}\mathcal{L}$\ \ where $\mathcal{D}_{4}\simeq
\mathbb{Z}_{4}$ is the 4 element discrete PCT group and $\mathcal{L}$
is the proper orthochronous Lorentz group. 
\begin{equation}
\mathcal{L}\subset \mathcal{S}\mathcal{O}( 1,3) \subset \mathcal{O}(
1,3) .
\end{equation}

The projective representations of $\mathcal{G}=\mathcal{O}( 1,3)
\otimes _{s}\mathcal{T}( 4) $ are equivalent to the unitary representations
$\varrho $ of the central extension of this group (Appendix A) \cite{wigner,Weinberg1}.\ \ The
algebraic extension is trivial and therefore the central extension
$\check{\mathcal{G}}$ of $\mathcal{G}$ is the cover, $\check{\mathcal{G}}\simeq
\overline{\mathcal{G}}$.\ \ The cover of the discrete group is itself\ \ ${\overline{\mathcal{D}}}_{4}\simeq
\mathcal{D}_{4}$ and the Lorentz group has a 2-1 cover $\overline{\mathcal{L}
}=\mathcal{S}\mathcal{L}( 2,\mathbb{C}) $. Special relativity is
then the unitary, or antiunitary, representation of the (extended)
Poincar\'e group $\mathcal{P}=\check{\mathcal{G}}\simeq \mathcal{D}_{4}\otimes
_{s}\mathcal{S}\mathcal{L}( 2,\mathbb{C}) \otimes _{s}\mathcal{T}(
4) $\footnote{PCT is an approximate symmetry that is not always
applicable.}.\ \ The two Casimir invariant operators for the Poincar\'e
group are $C_{1}=\eta ^{a,b} P_{a}P_{b}$ and $C_{2}= \eta ^{a,b}W_{a}W_{b}$
with $W_{a}=\epsilon _{a}^{b,c,d}L_{b,c}P_{d}$.\ \ 

The unitary representations of the Poincar\'e group have been extensively
studied and it is not our purpose to repeat it here.\ \ We note
only that the usual single particle wave equations for single particle
states, Klein-Gordon, Dirac, Maxwell and so forth result\ \ from
the\ \ solution of the eigenvalue equations of the Hermitian representation
of the Casimir invariants 
\begin{equation}
\left. {\hat{C}}_{\alpha }\ \ \left| \psi \right. \right\rangle
=c_{\alpha }\left. \left| \psi \right. \right\rangle   \mathrm{with}
\left. \left| \psi \right. \right\rangle  \in {\text{\boldmath $\mathrm{H}$}}^{\varrho
},\ \ \ \alpha =1,2...N_{c}.%
\label{field equation definition}
\end{equation}

\noindent where the eigenvalues are $c_{1}=-\mu ^{2}$ and $c_{2}=s(
s+1) \mu ^{2}$ with $\mu $ interpreted as mass and $s$ as intrinsic
spin or helicity. 

This method may be applied to other homogeneous relativity groups
$ \mathcal{K}$. The relativistic quantum mechanics is the projective
representations of the inhomogeneous group $\mathcal{G}=\mathcal{K}\otimes
_{s}\mathcal{T}( n) $.\ \ Projective representations of a group
$\mathcal{G}$ are equivalent to unitary representations of the central
extension of the group, $\check{\mathcal{G}}$ \cite{Weinberg1}.\ \ \ \ The
unitary representations, and the corresponding Hilbert space of
states, are determined by the Mackey method. The single particle
wave equations are given by the eigenvalue equations of the representations
of the Casimir invariant operators (47)\ \ \cite{Low}. 
\subsection{Reciprocally relativistic quantum mechanics.}

Reciprocally relativistic quantum mechanics generalizes special
relativistic quantum mechanics to noninertial frames. The method
directly follows the approach described in the previous section.\ \ Reciprocally
relativistic quantum mechanics is the projective representations
of the inhomogeneous unitary group. 

As in the special relativistic case, the quantum theory considers
the full symmetries that, in this case are the $\mathcal{U}( 1,3)
$ corresponding to the $\mathcal{O}( 2,6) $ invariance of the Born-Green
metric\footnote{It may be necessary to consider additional discrete
symmetries as in the special relativity case}. 
\subsubsection{Central extension: the quaplectic group}

We determine the central extension of the Inhomogeneous unitary
group in Appendix A.\ \ The result is that the central extension
of the group $\mathcal{U}( 1,3) \otimes _{s}\mathcal{T}( 8) $ is
the universal cover $\overline{\mathcal{Q}}( 1,3) $ of the quaplectic
group $\mathcal{Q}( 1,3) =\mathcal{U}( 1,3) \otimes _{s}\mathcal{H}(
4) $. The cover is 
\begin{equation}
 \overline{\mathcal{Q}}( 1,3) \simeq \mathcal{T}( 1) \otimes _{s}\mathcal{S}\mathcal{U}(
1,3) \otimes _{s}\mathcal{H}( 4) .
\end{equation}

Thus, using the same method as in special relativistic quantum mechanics,\ \ reciprocal
relativistic quantum mechanics is given in terms of\ \ the unitary
representations of $\overline{\mathcal{Q}}( 1,3) $\footnote{The
algebra of $\mathcal{Q}( 1,3) $ and its cover are the same and we
move between these relatively freely in these considerations. See
comment in Appendix C. }.\ \ 

An element $g$ of the special quaplectic group may be written as
realized as $g( \Gamma ,w,\iota ) =g( \Gamma ,0,0) \cdot g( I,w,\iota
) $ where $g( \Gamma ,0,0) \in \mathcal{S}\mathcal{U}( 1,3) $ and
$g( I,w,\iota ) \in \mathcal{H}( 4) $.\ \ This may be realized as
the real $10\times 10$ matrices
\begin{equation}
g( \Gamma ,w,\iota ) \simeq \left( \begin{array}{lll}
 \Gamma  & 0 & 0 \\
 0 & 1 & 0 \\
 0 & 0 & 1
\end{array}\right) \left( \begin{array}{lll}
 I & 0 & w \\
 \zeta \cdot w & 1 & \iota  \\
 0 & 0 & 1
\end{array}\right) .
\end{equation}

The $\Gamma $ are the homogeneous transformations $\Gamma ( v,f,r)
$ in $3+1$ dimensions between non-inertial frames that are defined
in (31).\ \ $w\in \mathbb{R}^{8}$ and $\iota \in \mathbb{R}$ parameterize
the Heisenberg group \cite{Low5}.\ \ Calculations are more convenient
if we choose a complex parameterization of this real group.\ \ \ In
natural units with $c=b=\hbar =1$, invariance of the orthogonal
and symplectic metrics requires that\ \ 
\begin{equation}
\Gamma =\left( \begin{array}{ll}
 \Lambda  & M \\
 -M & \Lambda 
\end{array}\right) 
\end{equation}

\noindent where $\Lambda \in \mathcal{S}\mathcal{O}( 1,3) $ is the
Lorentz subgroup.\ \ Then, with $\{w\}=\{x,y\}$, $x,y\in \mathbb{R}^{4}$
and $\{x\}=\{t,q\}$, $\{y\}=\{e,p\}$\footnote{Note that this means
the order of the coordinates (and basis) is now $\{t,q,e,p\}$ rather
than $\{t,q,p,e\}$. This simply a matter that the latter is preferable
for the introductory comments whereas this ordering enables the
complex basis to be most simply introduced. } 
\begin{equation}
\Xi =\frac{1}{2} \left( M+i \Lambda \right) ,\ \ \ \ z= \frac{1}{\sqrt{2}}\left(
x+i y\right) .
\end{equation}

\noindent $\Xi \in \mathcal{S}\mathcal{U}( 1,3) $ are $4\times 4$
complex matrices with unit determinant .\ \ If $\omega \in \mathcal{U}(
1) $ is a phase, then $\Upsilon =\omega \Xi $ is an element of\ \ the
full $\mathcal{U}( 1,3) =\mathcal{U}( 1) \otimes \mathcal{S}\mathcal{U}(
1,3) $ group (29). This may be written compactly as the complex
$6\times 6$ complex matrix realization of the quaplectic group 
\begin{equation}
g( \Upsilon ,z,\iota ) \simeq \left( \begin{array}{lll}
 \Upsilon  & 0 & \Upsilon \cdot z \\
 \overline{z} & 1 & \iota  \\
 0 & 0 & 1
\end{array}\right) .
\end{equation}

The algebra of the quaplectic group is 
\begin{equation}
\begin{array}{ll}
 \left[ A_{a,b},A_{c,d}\right]  =i(  \eta _{a,d}A_{c,b} -\eta _{b,c}A_{a,d}
) , & \left[ Z_{a}^{+},Z_{b}^{-}\right] =i \eta _{a,b}\mathit{I}\mathit{,}
\\
 \left[ A_{a,b},Z_{c}^{+}\right] =-i \eta _{a,c}Z_{b}^{+}, & \left[
A_{a,b},Z_{c}^{-}\right] =i \eta _{b,c}Z_{a}^{-},
\end{array}%
\label{Canonical Algebra}
\end{equation}

\noindent where $U=\eta ^{a,b}A_{a,b}$ is the generator of the $\mathcal{U}(
1) $ group. The Casimir invariants for the group are \cite{quesne}
\begin{equation}
\begin{array}{l}
 C_{0}= I,\ \ \ \ \ \ C_{1}=\eta ^{a,b}W_{a,b} ,\ \ \ \ \ ...  \\
 C_{4}=\eta ^{a,b}\eta ^{c,d}\eta ^{e,f}\eta ^{g,h}W_{h,a}W_{b,c}W_{d,e}W_{f,g},
\end{array}%
\label{Casimir invariants of cannonical group in complex basis}
\end{equation}

\noindent where it is noted that the number of independent Casimir
operators is 5 with\footnote{$Z_{b}^{-}Z_{a}^{+}$ may also be used
in this definition, or any linear combination with\ \ $Z_{a}^{+}Z_{b}^{-}$.\ \ \ From
the commutation relations, they differ only by a central element
$I$ that does not affect the definition of the Casimir invariant.
}\ \ 
\begin{equation}
W_{a,b}\doteq Z_{a}^{+}Z_{b}^{-}-I A_{a,b}.%
\label{W definition}
\end{equation}

\noindent Consequently, the second order invariant is of the form
\begin{equation}
C_{2}= N-I U%
\label{canonical algebra casimir second order simple form}
\end{equation}

\noindent where $N=\eta ^{a,b}{Z^{+}}_{a}{Z^{-}}_{b}$ and $U$ is
the generator of the algebra of $ \mathcal{U}$(1) defined above.
The commutation relations for the $W_{a,b}$ are 
\begin{equation}
\begin{array}{l}
 \left[ A_{a,b},W_{c,d}\right]  =i( \eta _{a,d}W_{b,c}-\eta _{b,c}W_{d,a})
, \\
  \left[ Z_{c}^{\pm },W_{a,b}\right] =0 ,
\end{array}%
\label{W commutators}
\end{equation}

\noindent and therefore $W_{c,d}$ are invariant under Weyl-Heisenberg
{\itshape translations}\footnote{The Weyl-Heisenberg group is the
semidirect product of two translation groups. In a sense, it is
a direct nonabelian generalization of our usual concept of translation.}.
It is important to note that both of the terms in $W_{a,b}$ are
required in order for the commutator with $Z_{c}^{\pm }$\ \ to vanish.
The $W_{c,d}$ obey the same commutation relations with $A_{a,b}$\ \ as
does $A_{c,d}$ in (53).\ \ The Casimir invariants of $\mathcal{U}(
1,n) $ are \cite{popov}
\begin{equation}
\begin{array}{l}
 D_{1}=\eta ^{a,b}A_{a,b} ,\ \ \ \ \ ...  \\
 D_{4}=\eta ^{a,b}\eta ^{c,d}\eta ^{e,f}\eta ^{g,h}A_{h,a}A_{b,c}A_{d,e}A_{f,g}.
\end{array}%
\label{Casimir invariants of unitary group in complex basis}
\end{equation}

\noindent Therefore, (53) are invariant under $\mathcal{U}( 1,n)
$ rotations and it follows that the $C_{\alpha }$ in (54) are Casimir
invariants of $\mathcal{Q}( 1,n) $.\ \ Note also that it follows
immediately that\ \ 
\begin{equation}
\left[ D_{\alpha },D_{\beta }\right] =0, \text{$[D_{\alpha },C_{\beta
}]=0$},\ \ \ \left[ C_{\alpha },C_{\beta }\right] =0 .%
\label{Quaplectic commute}
\end{equation}
\subsubsection{Unitary representations of the quapletic group: Hermitian
representations of its algebra}

The problem now is to determine the unitary representations of the
group and the corresponding Hermitian representation of the algebra.\ \ This
is a semidirect product with an nonabelian normal subgroup for which
the Mackey representation theory is applicable (see Appendix B)
\cite{Low,Low2}.\ \ 

The results are as follows.\ \ There are two classes of representations
corresponding to whether the eigenvalues of the representation of
$C_{0}=I$ are zero or non-zero. If $\hat{I}|\psi \rangle =0$, the
${Z^{\pm }}_{b}$ commute and this reduces to the degenerate case
of the inhomogeneous group where the normal subgroup is the abelian
translation group.\ \ This is not of further interest. 

If $\hat{I}|\psi \rangle \neq 0$, the little group is $\mathcal{U}(
1,3) $ itself and the stabilizer is the full quaplectic group. Thus,
the representations may be determined without requiring induction
to the full group from the stabilizer.\ \ 

Using the Mackey method, the unitary representation is $\varrho
=\sigma \otimes \rho $\ \ where $\sigma $ is a unitary representation
of the little group, which is $\mathcal{U}( 1,3) $, that acts on
a Hilbert space ${\text{\boldmath $\mathrm{H}$}}^{\sigma }$ and
$\rho $ is a projective representation of $\mathcal{Q}( 1,3) $ that
acts on the Hilbert space of the unitary representations $\xi $
of the normal subgroup $\mathcal{H}( 4) $. 

The unitary representations of\ \ $\mathcal{U}( 1,3) $ are known
and act on a countably infinite complex vector space ${\text{\boldmath
$\mathrm{H}$}}^{\sigma }\simeq \mathbb{V}^{\infty }$ for this non
compact case\ \ 
\begin{equation}
{\hat{\varepsilon }}_{a,b}= \sigma '\left( A_{a,b}\right) ,\ \ \ \ \ \ \sigma
'\left( Z_{a}^{\pm }\right) =0.
\end{equation}

\noindent The generators of the Hermitian representation of the
algebra of $\mathcal{U}( 1,3) $ have commutation relations 
\begin{equation}
\left[ {\hat{\varepsilon }}_{a,b},{\hat{\varepsilon }}_{c,d}\right]
=\eta _{b,c}{\hat{\varepsilon }}_{a,d}-\eta _{a,d}{\hat{\varepsilon
}}_{c,b}.
\end{equation}

The projective representation $\rho $ is an extension of $\xi $,
so that $\rho $ restricted to $\mathcal{H}( 4) $ is $\xi $.\ \ \ As
the Weyl-Heisenberg group is the semidirect product $\mathcal{H}(
n) =\mathcal{T}( n) \otimes _{s}\mathcal{T}( n+1) $, its unitary
representations $\xi $ may be determined using the Mackey method
\cite{Major,Low}. The Hilbert space is\ \ ${\text{\boldmath $\mathrm{H}$}}^{\xi
}\simeq {\text{\boldmath $\mathrm{L}$}}^{2}( \mathbb{R}^{4},\mathbb{C})
$. The representation $\xi $ of the Weyl-Heisenberg group is lifted
to the algebra to define ${\hat{Z}}_{a}^{\pm }=\varrho '(Z_{a}^{\pm
})=\xi '(Z_{a}^{\pm })$\ \ where, as usual, in a co-ordinate basis
\begin{equation}
\left\langle  x\right| {\hat{Z}}_{a}^{\pm }\left| \psi \right\rangle
=\left( x^{a}\pm \eta ^{a,b}\frac{\partial }{\partial x^{b}}\right)
\psi ( x) =\left( x^{a}\pm \frac{\partial }{\partial x_{b}}\right)
\psi ( x) .%
\label{coord rep of Z}
\end{equation}

These satisfy the algebra
\begin{equation}
\left[ {\hat{Z}}_{a}^{-},{\hat{Z}}_{b}^{+}\right] = \eta _{a,b}
\hat{I}=\eta _{a,b}
\end{equation}

\noindent where $ \hat{I}|\psi \rangle =|\psi \rangle $ and therefore
the Casimir eigenvalue $c_{0}=1$\footnote{If $c_{0}\neq 1$, then
${\hat{W}}_{a,b}=(1-{c}_{0}){\hat{Z}}_{a}^{+}{\hat{Z}}_{b}^{-} -
{\hat{\varepsilon }}_{a,b}$ and this does not commute with the ${\hat{Z}}_{a}^{\pm
}$ and therefore cannot be used to construct the Casimirs.\ \ This
is incorrect in\ \ [12]. Thanks to P. Jarvis for the correct solution.}.\ \ 

As $\mathcal{H}( 4) $ is nonabelian, it is necessary to construct
a projective representation $\rho $\ \ that reduces to\ \ $\xi $
when restricted to the normal subgroup, $\rho |_{\mathcal{H}( 4)
}=\xi $.\ \ $\rho  $ acts on the Hilbert space ${\text{\boldmath
$\mathrm{H}$}}^{\xi }$.\ \ \ This is equivalent to determining a
Hermitian\footnote{The quaplectic group is its own algebraic central
extension. } representation $\ \ \rho '$ of the algebra acting on
${\text{\boldmath $\mathrm{H}$}}^{\xi }$.\ \ \ As shown in Appendix
C, this extension is given by 
\begin{equation}
{\hat{Z}}_{a,b }=\ \ \rho '\left( A_{a,b}\right) =\xi '\left( Z_{a}^{+}\right)
\cdot \xi '\left( Z_{b}^{-}\right) = {\hat{Z}}_{a}^{+}{\hat{Z}}_{b}^{-}.%
\label{Zab definition}
\end{equation}

\noindent These Hermitian differential operators satisfy the commutation
relations\ \ 
\begin{equation}
\begin{array}{l}
 \left[ {\hat{Z}}_{a,b},{\hat{Z}}_{c}^{+}\right] =-\eta _{a,c}\ \ {\hat{Z}}_{b}^{+},\ \ \left[
{\hat{Z}}_{a,b},{\hat{Z}}_{c}^{-}\right] =\eta _{b,c}\ \ {\hat{Z}}_{a}^{-},
\\
 \left[ {\hat{Z}}_{a,b},{\hat{Z}}_{c,d}\right] =\eta _{b,c}{\hat{Z}}_{a,d}-\eta
_{a,d}{\hat{Z}}_{c,b} .
\end{array}%
\label{quaplectic representation commutation relations}
\end{equation}

\noindent with ${\hat{Z}}_{a,b}$ and ${\hat{Z}}_{c}^{\pm }$ commuting
with the ${\hat{\varepsilon }}_{c,b}$.\ \ 

Finally, as $\varrho '=\sigma '\oplus \rho '$ act on the Hilbert
space ${\text{\boldmath $\mathrm{H}$}}^{\varrho }\simeq \mathbb{V}^{\infty
}\otimes {\text{\boldmath $\mathrm{L}$}}^{2}( \mathbb{R}^{4},\mathbb{C})
$ where the generators are given by\ \ $\varrho '(A_{a,b})={\hat{A}}_{a,b}={\hat{Z}}_{a,b}+{\hat{\varepsilon
}}_{a,b}$.\ \ The ${\hat{W}}_{a,b}= \varrho ( W_{a,b}) $ defined
in (55), that are used in the definition of the Casimir invariants,
are
\begin{equation}
{\hat{W}}_{a,b}={\hat{Z}}_{a}^{+} {\hat{Z}}_{b}^{-}-\hat{ I} {\hat{A}}_{a,b}=
- {\hat{\varepsilon }}_{a,b} .
\end{equation}

\noindent The Casimir eigenvalue equations in (47) may then be written
out explicitly as 
\begin{equation}
\begin{array}{l}
 \left. {\hat{C}}_{0}\ \ \left| \psi \right. \right\rangle  =\left.
\left. \hat{I}\ \ \left| \psi \right. \right\rangle  =\left| \psi
\right. \right\rangle  ,\ \ \ \ \ \ \  \\
 \left. {\hat{C}}_{1}\ \ \left| \psi \right. \right\rangle  =\eta
^{a,b}{\hat{W}}_{a,b} \left. \left| \psi \right. \right\rangle 
=\left. \left. -\eta ^{a,b}{\hat{\varepsilon }}_{a,b}\left| \psi
\right. \right\rangle   =c_{1 }\left| \psi \right. \right\rangle
,\ \ \ \ \ ... \\
 \left. {\hat{C}}_{4}\ \ \left| \psi \right. \right\rangle  =\left.
\eta ^{a,b}\eta ^{c,d}\eta ^{e,f}\eta ^{g,h}{\hat{\varepsilon }}_{h,a}{\hat{\varepsilon
}}_{b,c}{\hat{\varepsilon }}_{d,e}{\hat{\varepsilon }}_{f,g} \left|
\psi \right. \right\rangle  =c_{4 }\left. \left| \psi \right. \right\rangle
.
\end{array}%
\label{Rep of Casimir invariants of cannonical group in complex
basis}
\end{equation}

\noindent The $c_{\alpha }$ label irreducible representations\footnote{The
Casimir invariants are constant for each irreducible representation.
However, they may not form a complete set, additional labels may
be required to completely specify the irreducible representations.
For semisimple groups and the Poincar\'e group, they are sufficient.
}and are given in terms of the Casimir invariants of the group $\mathcal{U}(
1,3) $ through the $\sigma $ representation. 

The Casimir invariant operators ${\hat{D}}_{\alpha }=\varrho '(D_{\alpha
})$ of the unitary group are \cite{popov} 
\begin{equation}
\left. {\hat{D}}_{\alpha }\left| \psi \right. \right\rangle  =d_{\alpha
} \left. \left| \psi \right. \right\rangle   \mathrm{with}  \left.
\left| \psi \right. \right\rangle  \in {\text{\boldmath $\mathrm{H}$}}^{\varrho
},\ \ \ \alpha =1,2...N_{u}=n.%
\label{field equation definition}
\end{equation}

\noindent where the representations of the Casimir operators are
\begin{equation}
\begin{array}{l}
 \left. {\hat{D}}_{1}\left| \psi \right. \right\rangle  =\eta ^{a,b}{\hat{A}}_{a,b}\left.
\left| \psi \right. \right\rangle  =\left. d_{1} \left| \psi \right.
\right\rangle   ,\ \ \ \ \ ...  \\
 \left. {\hat{D}}_{4}\left| \psi \right. \right\rangle  =\left.
\left. \eta ^{a,b}\eta ^{c,d}\eta ^{e,f}\eta ^{g,h}{\hat{A}}_{h,a}{\hat{A}}_{b,c}{\hat{A}}_{d,e}{\hat{A}}_{f,g}\left|
\psi \right. \right\rangle  =d_{4} \left| \psi \right. \right\rangle
,
\end{array}%
\label{Rep of Casimir invariants of unitary group in complex basis}
\end{equation}

\noindent and where\ \ ${\hat{A}}_{a,b}={\hat{Z}}_{a,b}+{\hat{\varepsilon
}}_{a,b}$.\ \ \ Substituting these into the expressions and simplifying
(see Appendix C) results in the equations
\begin{equation}
\begin{array}{l}
 \eta _{a,b}( x^{a}+\frac{\partial }{\partial x_{a}}) \left( x^{b}-\frac{\partial
}{\partial x_{b}}\right) =f_{1}( c_{1},d_{1}) \psi ( x) , \\
 {\hat{\varepsilon }}_{b,a}( x^{a}+\frac{\partial }{\partial x_{a}})
\left( x^{b}-\frac{\partial }{\partial x_{b}}\right) \psi ( x) =f_{2}(
c_{1},c_{2},d_{1},d_{2}) \psi ( x) , \\
 {\hat{\varepsilon }}_{b}^{c}{\hat{\varepsilon }}_{c,a}( x^{a}+\frac{\partial
}{\partial x_{a}}) \left( x^{b}-\frac{\partial }{\partial x_{b}}\right)
\psi ( x) =f_{3}( c_{1},...c_{3},d_{1}...d_{3}) \psi ( x) , \\
 {\hat{\varepsilon }}_{b}^{d}{\hat{\varepsilon }}_{d}^{c}{\hat{\varepsilon
}}_{c,a}( x^{a}+\frac{\partial }{\partial x_{a}}) \left( x^{b}-\frac{\partial
}{\partial x_{b}}\right) \psi ( x) =f_{4}( c_{1},...c_{4},d_{1},...d_{4})
\psi ( x) ,
\end{array}%
\label{quaplectic wave equations}
\end{equation}

\noindent with ${\hat{\varepsilon }}_{b}^{c}=\eta ^{c,a}{\hat{\varepsilon
}}_{a,b}$. The $f_{\alpha }( c_{1},...,d_{1},...) $ are polynomials
in $c_{\beta }$ and $d_{\beta }$.\ \ The $c_{\beta }$ label irreducible
representations and the $d_{\beta }$ label states in the irreducible
representations\footnote{As noted previously, this labeling may
not be complete}.\ \ This is the set of wave equations\ \ that results
from one new physical assumption, the Born-Green metric (27).\ \ The
${\hat{\varepsilon }}_{c,d}$ are countably infinite dimensional
matrices as they are the Hermitian representations of the algebra
of $\mathcal{U}( 1,3) $. The wave functions are functions of $\mathbb{R}^{4}$
and have countably infinite number of components.\ \ The Hilbert
space on which these act is ${\text{\boldmath $\mathrm{H}$}}^{\varrho
}\simeq \mathbb{V}^{\infty }\otimes {\text{\boldmath $\mathrm{L}$}}^{2}(
\mathbb{R}^{4},\mathbb{C}) $.\ \ The wave functions are elements
of ${\text{\boldmath $\mathrm{L}$}}^{2}( \mathbb{R}^{4}) $ and not
${\text{\boldmath $\mathrm{L}$}}^{2}( \mathbb{R}^{8}) $ because
the Weyl-Heisenberg group is required by the central extension as
a direct consequence of requiring projective representations. There
is no need for a separate\ \ {\itshape quantization }procedure.
\section{Discussion}

The theory outlined seeks to generalize special relativity to noninertial
frames\footnote{Note that in general relativity, particles that
are only under the influence of gravity follow geodesics and so
are locally inertial. The problem of determining a corresponding
generalization of the theory described here to a manifold that is
curved (and in this case noncommutative) has not yet been studied.
}. This introduces a reciprocally dual\ \ relativity principle that
requires forces between particle states to be bounded by a universal
constant $b$ in addition to special relativity requiring velocity
between particle states to be bounded by $c$. Both velocity and
force are relative. There is no longer the concept of an absolute
inertial frame nor an absolute rest frame

The theory may be regarded as a higher dimensional {\itshape spacetime
}where the additional dimensions of this higher dimensional {\itshape
spacetime} are energy and momentum.\ \ That is, these additional
dimensions are just as {\itshape physical} as the position and time
degrees of freedom. Position-time space\footnote{That is, our usual
concept of {\itshape spacetime}} is not an invariant subspace for
noninertial observers for which the transformations between frames
are given by the unitary group. Different noninertial observers
define this subspace differently. In this sense, position-time space
(that is, our current concept of spacetime) has become relative.\ \ Energy-momentum
may transform into space-time. In the inertial limit $b\rightarrow
\infty $, an invariant position-time space is recovered \cite{jarvis}.

The bound $b$ of relative forces means that force singularities
cannot exist.\ \ A theory that is invariant under the noninertial
group will bound these effects through noninertial relativistic
effects that result from the generalized concept of contractions
and dilations (34). Other approaches to resolving these singularities\ \ include
assume a minimum length or a maximum acceleration \cite{Schuller-1,Jacobson,smolin,cailiello}.

Quantum mechanics is formulated by identifying physical states with
rays in a Hilbert space. This leads to projective representations
of the inhomogeneous noninertial relativity group. These are the
unitary representations of the central extension that is the cover
of the quaplectic group. This group has the Weyl-Heisenberg group
as the normal subgroup with the associated Heisenberg algebra.\ \ Thus
the basic Heisenberg relations result from requiring projective
representations of the inhomogeneous group for the noninertial frames.

The Hilbert space of the Weyl-Heisenberg group $\mathcal{H}( 4)
$ is ${\text{\boldmath $\mathrm{L}$}}^{2}( \mathbb{R}^{4}) $ not
${\text{\boldmath $\mathrm{L}$}}^{2}( \mathbb{R}^{8}) $.\ \ Thus
the wave functions are a function of a four dimensional subspace
of\ \ commuting degrees of freedom. One such set is position-time
but three additional canonical sets may also be used.\ \ This is
simply represented by arranging the basis of the algebra $\{T,Q_{i},P_{i},E\}$\ \ into
a quad with the four generators on each face commuting.
\begin{equation}
\begin{array}{ll}
 T & Q_{i} \\
 P_{i} & E
\end{array}
\end{equation}

Just as nonrelativistic mechanics must be obtained from the limit
$c\rightarrow \infty $, we must obtain special relativistic quantum
mechanics from the reciprocal relativistic quantum mechanics in
the limit $b\rightarrow \infty $. The next step in this investigation
is to determine whether the wave equations for single particle states
determined from the representations of the quaplectic group\ \ reduce
to the special relativity inertial case in the limit $b\rightarrow
\infty $\footnote{At first glance this does not seem likely to be
the case particularly due to the\ \ wave functions having countably
infinite dimensional components.\ \ For an example of the effects
of limits, recall that the ordinary three dimensional nonrelativistic
quantum harmonic oscillator has the Hilbert space\ \ $\mathbb{Z}\otimes
{\text{\boldmath $\mathrm{L}$}}^{2}( \mathbb{R}^{3},\mathbb{C})
$. The classical limit has a Hilbert space ${\text{\boldmath $\mathrm{L}$}}^{2}(
\mathbb{S}^{2},\mathbb{C}) $. }. 

I would like to thank Peter Jarvis for discussions on the ideas
presented here that have greatly facilitated their development.
\section{Appendix A: Central extensions\ \ of Lie groups}\label{Appendix
central extension}

The projective representation of a Lie group is equivalent the unitary
(or antiunitary) representation of the central extension of the
group \cite{Weinberg1}. The central extension is algebraic or topological
or both. Consider first the algebraic extension.\ \ Suppose $\{X_{\alpha
}\}$ are the generators of the Lie group $\mathcal{G}$ with commutators
\begin{equation}
\left[ X_{\alpha },X_{\beta }\right] =C_{\alpha ,\beta }^{\gamma
}X_{\gamma }
\end{equation}

\noindent with $\alpha ,\beta ,...=1,..\dim ( \mathcal{G}) $.\ \ Then,
the central extension is the addition of a central generator $I_{a}$,\ \ $[X_{\alpha
},I_{a}]=0$,\ \ with commutator
\begin{equation}
\left[ X_{\alpha },X_{\beta }\right] =C_{\alpha ,\beta }^{\gamma
}X_{\gamma }+{\tilde{C}}_{\alpha ,\beta }^{a}I_{a}.
\end{equation}

\noindent $a=1,...m$\ \ where $m$ is the dimension of the central
extension. This new commutation relation must also satisfy the Jacobi
identities
\begin{equation}
\left[ X_{\gamma },\left[ X_{\alpha },X_{\beta }\right] \right]
+\left[ X_{\alpha },\left[ X_{\beta },X_{\gamma }\right] \right]
+\left[ X_{\beta },\left[ X_{\gamma },X_{\alpha }\right] \right]
=0.
\end{equation}

\noindent Clearly ${\tilde{C}}_{\alpha ,\beta }^{\gamma }=C_{\alpha
,\beta }^{\gamma }$ is a trivial solution involving only the redefinition
$X_{\gamma }\rightarrow X_{\gamma }+I_{\gamma }$ and need not be
considered. 

A direct calculation using the structure constants for the inhomogeneous
Lorentz group shows that there are no nontrivial solutions in this
case.\ \ On the other hand, a direct computation for the inhomogeneous
unitary group,\ \ $\mathcal{U}( 1,n) \otimes _{s}\mathcal{T}( 2n+2)
$, with the generators of the algebra satisfying (53) shows that
there is a one dimensional, ($m=1$) central extension such that
$[Z_{a}^{+},Z_{b}^{-}]=i I \eta _{a,b}$.\ \ This is the algebra
of the Heisenberg group.\footnote{A semidirect product must be a
subgroup of the group of automorphisms of the normal subgroup. For
the Heisenberg group $\mathcal{H}( n) $, this is essentially $\mathcal{S}p(
2n) \otimes _{s}\mathcal{H}( n) $. $\operatorname{S\mathcal{O}}(
1,3) $ is not a subgroup of\ \ $\mathcal{S}p( 4) $ and so this central
extension is not possible whereas $\mathcal{U}( 1,3) \subset \mathcal{S}p(
8) $ and so it is possible in this case. }\ \ 

The topological extension is the universal cover $\overline{\mathcal{G}}$\ \ of
$\mathcal{G}$ where
\begin{equation}
\pi :\overline{\mathcal{G}}\rightarrow \mathcal{G}\ \ \ \mathrm{with}
\ker ( \pi ) \simeq \mathcal{D}\text{}.
\end{equation}

\noindent $ \mathcal{D}$ is an abelian discrete subgroup that is
the central extension that is also the first homotopy group.\ \ For
the case of a semidirect product of matrix groups where $\mathcal{G}=\mathcal{K}\otimes
_{s}\mathcal{N}$, the cover is given by $\overline{\mathcal{G}}=\overline{\mathcal{K}}\otimes
_{s}\overline{\mathcal{N}}$.

It is well known that $\mathcal{S}\mathcal{L}(2,\mathbb{C}$)\ \ is
a double cover of the Lorentz group $\mathcal{L}$, $\pi :\mathcal{S}\mathcal{L}(
2,\mathbb{C}) \rightarrow \mathcal{L}\ \ \ \mathrm{with} \ker (
\pi ) \simeq \mathcal{D}\simeq \mathbb{Z}_{2}\text{}$. The translation
group is simply connected and is its own cover and so the central
extension of the inhomogeneous proper orthochronous Lorentz group
is $\mathcal{S}\mathcal{L}( 2,\mathbb{C}) \otimes _{s}\mathcal{T}(
4) $.

The quaplectic group may be written as $\mathcal{Q}( 1,3) =\mathcal{U}(
1) \otimes _{s}\mathcal{S}\mathcal{U}( 1,3) \otimes _{s}\mathcal{H}(
4) $.\ \ \ $\mathcal{S}\mathcal{U}( 1,3) $ is simply connected and
is its own cover. The cover of the\ \ $\mathcal{U}( 1) $ group is
the translation group
\begin{equation}
\pi :\mathcal{T}( 1) \rightarrow \mathcal{U}( 1) \ \ \ \mathrm{with}
\ker ( \pi ) \simeq \mathbb{Z}\text{}.
\end{equation}

\noindent The Weyl-Heisenberg group is simply connected and is its
own cover. Therefore, the central extension of the inhomogeneous
unitary group\ \ $\mathcal{U}( 1,3) \otimes _{s}\mathcal{T}( 8)
$ is 
\begin{equation}
\overline{\mathcal{Q}}( 1,3) \simeq \mathcal{T}( 1) \otimes _{s}\mathcal{S}\mathcal{U}(
1,3) \otimes _{s}\mathcal{H}( 4) .
\end{equation}
\section{Appendix B: Unitary irreducible representations of\ \ semidirect
product groups }\label{Appendix mackey}

The problem of determining the unitary irreducible representations
of a general class of semidirect product groups has been solved
by Mackey \cite{mackey}. Application to the Weyl-Heisenberg and
quaplectic groups may be found in \cite{Major,Wolf}.\ \ Mackey formulates
the theorem for a very general class of groups. All the groups under
consideration are well behaved, real matrix Lie groups and their
covers for which the conditions are sufficient for the theorems
to apply.\ \ \ The Mackey theorems are reviewed in \cite{Low} and
briefly summarized here. In addition, the manner in which the results
lift to the algebra is given as they are required for the determination
of the field equations. 
\subsection{{\bfseries Unitary irreducible representations of the
Lie group}}

Suppose that $\mathcal{N}$ and $ \mathcal{K}$\ \ are matrix groups
that are algebraic with unitary irreducible representations $\xi
$ and $\sigma $ on the respective Hilbert spaces ${\text{\boldmath
$\mathrm{H}$}}^{\xi }$ and ${\text{\boldmath $\mathrm{H}$}}^{\sigma
}$. Then for $z\in \mathcal{N}$,\ \ \ and $k\in \mathcal{K}$
\begin{equation}
\begin{array}{l}
 \xi ( z) :{\text{\boldmath $\mathrm{H}$}}^{\xi }\rightarrow {\text{\boldmath
$\mathrm{H}$}}^{\xi }:\left. \left| \phi \right. \right\rangle 
\mapsto \tilde{\left. \left| \phi \right. \right\rangle  }=\left.
\xi ( z) \left| \phi \right. \right\rangle  , \\
 \sigma ( k) :{\text{\boldmath $\mathrm{H}$}}^{\sigma }\rightarrow
{\text{\boldmath $\mathrm{H}$}}^{\sigma }:\left. \left| \varphi
\right. \right\rangle  \mapsto \tilde{\left. \left| \varphi \right.
\right\rangle  }=\sigma ( k) \left. \left| \varphi \right. \right\rangle
.
\end{array}
\end{equation}

The general problem is to determine the unitary irreducible representations
$\varrho $, and the Hilbert space ${\text{\boldmath $\mathrm{H}$}}^{\varrho
}$\ \ on which it acts, of the semidirect product $\mathcal{G}=\mathcal{K}\otimes
_{s}\mathcal{N}$,
\begin{equation}
\varrho ( g) :{\text{\boldmath $\mathrm{H}$}}^{\sigma }\rightarrow
{\text{\boldmath $\mathrm{H}$}}^{\sigma }:\left. \left| \psi \right.
\right\rangle  \mapsto \tilde{\left. \left| \psi \right. \right\rangle
}=\varrho ( g) \left. \left| \psi \right. \right\rangle  .
\end{equation}

The Mackey theorems state that these unitary irreducible representations
$\varrho $ may be constructed by first determining the representations\ \ $\varrho
\mbox{}^{\circ}$ of the stabilizer groups, $\mathcal{G}\mbox{}^{\circ}\subseteq
\mathcal{G}$ and then using an induction theorem to obtain the representations
on the full group $\mathcal{G}$. A sufficient condition for the
Mackey method to apply is that $\mathcal{G}, \mathcal{K}$ and $\mathcal{N}$
are matrix groups that are algebraic in the sense that they are
defined by polynomial constraints on the general linear groups.

The stabilizer group is $\mathcal{G}\mbox{}^{\circ}=\mathcal{K}\mbox{}^{\circ}\otimes
_{s}\mathcal{N}$\ \ where the little group $\mathcal{K}\mbox{}^{\circ}\subset
\mathcal{K}$ is defined for each of the {\itshape orbits}. These
orbits are\ \ defined by the natural action of elements $k\in \mathcal{K}$
on the unitary dual $\hat{\mathcal{N}}$ of $\mathcal{N}$. The action
defining the orbits is $k:\hat{\mathcal{N}}\rightarrow \hat{\mathcal{N}}:\xi
\mapsto \tilde{\xi }=k \xi $\ \ where $(k \xi )(z)=\xi ( k\cdot
z\cdot k^{-1}) $ for all $a\in \mathcal{N}$. The little groups $
\mathcal{K}$$ \mbox{}^{\circ}$ are defined by a certain fixed point
condition on each these orbits. 

For the case that $\mathcal{N}$ is abelian, the fixed point condition
is $k \xi =\xi $ and the little group is $\mathcal{K}\mbox{}^{\circ}=\{k|k
\xi =\xi \}$. The representation $\varrho \mbox{}^{\circ}=\sigma
\otimes \chi $ acts on the Hilbert space ${\text{\boldmath $\mathrm{H}$}}^{\varrho
\mbox{}^{\circ}}\simeq {\text{\boldmath $\mathrm{H}$}}^{\sigma }\otimes
\mathbb{C}$.\ \ Note that as $\mathcal{N}$ is abelian, $\mathcal{N}\simeq
\mathbb{R}^{n}$ under addition and the representations are the characters\ \ $\xi
_{c}( z) =\chi _{c}( z) = e ^{i z\cdot c}$ and\ \ ${\text{\boldmath
$\mathrm{H}$}}^{\xi }\simeq \mathbb{C}.$

For the case that $\mathcal{N}$ is not abelian, the fixed point
condition is $k \xi =\rho ( k) \xi  {\rho ( k) }^{-1}$ and the representation
$\varrho \mbox{}^{\circ}=\sigma \otimes \rho $ acts on the Hilbert
space ${\text{\boldmath $\mathrm{H}$}}^{\varrho \mbox{}^{\circ}}\simeq
{\text{\boldmath $\mathrm{H}$}}^{\sigma }\otimes {\text{\boldmath
$\mathrm{H}$}}^{\xi }$.\ \ $\rho $ is a projective extension of
the representation $\xi $ to\ \ $\mathcal{G}\mbox{}^{\circ}$, $\rho
( g) :{\text{\boldmath $\mathrm{H}$}}^{\xi }\rightarrow {\text{\boldmath
$\mathrm{H}$}}^{\xi }$\ \ for $g\in \mathcal{G}\mbox{}^{\circ}$
with $\rho |_{\mathcal{N}}\simeq \xi $. If $\mathcal{N}$ is abelian,
the extension is trivial, $\rho |_{\mathcal{K}}\simeq 1$ and this
reduces to the abelian case above.\ \ Otherwise, the projective
representations $\rho $ are equivalent to the unitary representations
of the central extension $\check{\mathcal{G}}\mbox{}^{\circ}$ of
$\mathcal{G}\mbox{}^{\circ}$ using the method of Appendix A.

If the stabilizer is equal to the group, $\mathcal{G}=\mathcal{G}\mbox{}^{\circ}$
we are done. Otherwise the Mackey induction theorem is required
to induce the representation on the full group \cite{Low}. As the
induction theorem is not required for the quaplectic group, it is
not reviewed further here. 
\subsection{{\bfseries Hermitian representation of the Lie algebra}}

These unitary representations may be lifted to the algebra. Define
$T_{e}\xi =\xi ^{\prime }$,\ \ $T_{e}\sigma =\sigma ^{\prime }$\ \ and
$T_{e}\varrho \mbox{}^{\circ}={\varrho \mbox{}^{\circ}}^{\prime
}$. Assume that $\mathcal{G}\mbox{}^{\circ}$ is the central extension
so that representations of the group are unitary and the algebra
are Hermitian. Then for $Z\in \text{\boldmath $\mathrm{a}$}( \mathcal{N})
\simeq T_{e}\mathcal{N}$, $A\in \text{\boldmath $\mathrm{a}$}( \mathcal{K}\mbox{}^{\circ})
$ and $W=A+Z\in \text{\boldmath $\mathrm{a}$}( \mathcal{G}) $ we
have
\begin{equation}
\begin{array}{l}
 {\varrho \mbox{}^{\circ}}^{\prime }( W) :{\text{\boldmath $\mathrm{H}$}}^{\varrho
\mbox{}^{\circ}}\rightarrow {\text{\boldmath $\mathrm{H}$}}^{\varrho
\mbox{}^{\circ}} \\
 =\sigma ^{\prime }( A) \oplus \rho ^{\prime }( W) :{\text{\boldmath
$\mathrm{H}$}}^{\sigma }\otimes {\text{\boldmath $\mathrm{H}$}}^{\xi
}\rightarrow {\text{\boldmath $\mathrm{H}$}}^{\sigma }\otimes {\text{\boldmath
$\mathrm{H}$}}^{\xi } \\
 \left. \left. \left. :\left. \left| \psi \right. \right\rangle
\mapsto \tilde{\left. \left| \psi \right. \right\rangle  }=\sigma
^{\prime }( A) \ \ \left| \varphi \right. \right\rangle  \otimes
\left| \phi \right. \right\rangle  \oplus \left| \varphi \right.
\right\rangle  \otimes \rho ^{\prime }( W) \ \ \left. \left| \phi
\right. \right\rangle  .
\end{array}
\end{equation}

The basis of the algebra satisfies the Lie algebra 
\begin{equation}
\begin{array}{l}
 \left[ A_{\mu },A_{\nu }\right] =c_{\mu ,\nu }^{\lambda } A_{\lambda
}, \\
 \left[ Z_{\alpha },Z_{\beta }\right] =c_{\alpha ,\beta }^{\gamma
} Z_{\gamma }, \\
 \left[ A_{\mu },Z_{\alpha }\right] =c_{\mu ,\alpha }^{\nu } A_{\nu
}.
\end{array}
\end{equation}

\noindent where $\alpha ,\beta ..=1...\dim ( \mathcal{N}) $ and
$\mu ,\nu ..=1,..\dim ( \mathcal{K}) $. Then, the Hermitian\ \ $\rho$
of the generators (of the central extension) satisfies the commutation
relations 
\begin{equation}
\begin{array}{l}
 \left[ \rho ^{\prime }( A_{\mu }) ,\rho ^{\prime }( A_{\nu }) \right]
=i c_{\mu ,\nu }^{\lambda } \rho ^{\prime }( A_{\lambda }) , \\
 \left[ \rho ^{\prime }( Z_{\alpha }) ,\rho ^{\prime }( Z_{\beta
}) \right] =i c_{\alpha ,\beta }^{\gamma } \rho ^{\prime }( Z_{\gamma
}) , \\
 \left[ \rho ^{\prime }( A_{\mu }) ,\rho ^{\prime }( Z_{\alpha })
\right] =i c_{\mu ,\alpha }^{\nu } \rho ^{\prime }( A_{\nu }) .
\end{array}%
\label{projective rep commutators}
\end{equation}

As we are using Hermitian operators (instead of anti-Hermitian operators),
an $i$ appears in the exponential $\varrho ( k) =e^{-i\hat{A}}$,
$\varrho ( n) =e^{-i\hat{Z}}$.\ \ The $\rho ^{\prime }( A_{\alpha
}) $ act on the Hilbert space ${\text{\boldmath $\mathrm{H}$}}^{\xi
}$ and therefore must be elements of the enveloping algebra $\text{\boldmath
$\mathrm{e}$}( \mathcal{N}) \simeq \text{\boldmath $a$}( \mathcal{N})
\oplus \text{\boldmath $a$}( \mathcal{N}) \otimes \text{\boldmath
$a$}( \mathcal{N}) \oplus ...$ .\ \ Therefore 
\begin{equation}
{\rho  }^{\prime }\left( A_{\mu }\right) = d_{\mu }^{\alpha }{\xi
}^{\prime }\left( Z_{\alpha }\right) + d_{\mu }^{\alpha ,\beta }{\xi
}^{\prime }\left( Z_{\alpha }\right) {\xi  }^{\prime }\left( Z_{\beta
}\right) +....
\end{equation}

\noindent These may be substituted into the commutation relations
above to determine the constants $\{d_{\mu }^{\alpha },d_{\mu }^{\alpha
,\beta },...\}$.\ \ In particular for, the quaplectic group, this
leads to (64).
\section{Appendix C: Wave equations of the quaplectic group }\label{Appendix
calcs}

The eigenvalue equations of the Hermitian representations of the
Casimir invariant operators define the wave equations that are the
single particle state equations for the theory (47).\ \ These are
given explicitly in (67-69).\ \ In the following it is shown that\ \ these
reduce to (70).\ \ First note that, from\ \ (69)
\begin{equation}
\eta ^{a,b}{\hat{A}}_{a,b}\left. \left| \psi \right. \right\rangle
=\eta ^{a,b}( {\hat{Z}}_{a,b}\mathrm{+}{\hat{\varepsilon }}_{a,b})
\left. \left| \psi \right. \right\rangle  =d_{1} \left. \left| \psi
\right. \right\rangle   
\end{equation}

\noindent and therefore 
\begin{equation}
\hat{N}\left| \psi \right\rangle  =\eta ^{a,b}{\hat{Z}}_{a}^{+}{\hat{Z}}_{b}^{-}
\left| \psi \right\rangle  =f_{1}( c_{1},d_{1})  \left| \psi \right\rangle
\label{Rep of Casimir invariants of unitary group in complex basis}
\end{equation}

\noindent with $f_{1}( c_{1},d_{1}) =(d_{1}+c_{1})$.\ \ As both
$c_{1}$ and $d_{1}$ are Casimir invariant constants of the $\mathcal{T}(
1) $ group, $d_{1},c_{1}\in \mathbb{R}$.\ \ (Note that for $\mathcal{U}(
1) $ in the\ \ $\mathcal{Q}( 1,3) $ case, $d_{1},c_{1}\in \mathbb{N}$.)\ \ \ In
a coordinate basis (62), this is the relativistic oscillator\ \ \ 
\[
\left( \frac{\partial ^{2}}{\partial t^{2}}- \frac{\partial ^{2}}{\partial
q^{2}}-t^{2}+q^{2}-2m-c_{1}-n+1\right) \psi ( t,q) =0 
\]

\noindent where we set $d_{1}=2m$. Natural units $c=b=\hbar =1$
are being used.\ \ Boundary conditions that the wave function vanishes
at infinity require that $c_{1}=0$ and $m\in \mathbb{N}$.\ \ \ It
is important to emphasize that $\psi ( t,q) \in {\text{\boldmath
$\mathrm{L}$}}^{2}( \mathbb{R}^{4},\mathbb{C}) $ and so is not constrained
to a mass shell that causes problems in the interpretation of this
equation in the context of special relativistic quantum mechanics.

The next equation is\ \ 
\begin{equation}
\eta ^{b,c}\eta ^{a,d}{\hat{A}}_{a,b}{\hat{A}}_{c,d}\left. \left|
\psi \right. \right\rangle  =\eta ^{b,c}\eta ^{a,d}( {\hat{Z}}_{a,b}\mathrm{+}{\hat{\varepsilon
}}_{a,b}) \left( {\hat{Z}}_{c,d}\mathrm{+}{\hat{\varepsilon }}_{c,d}\right)
\left. \left| \psi \right. \right\rangle  =d_{1} \left. \left| \psi
\right. \right\rangle   %
\label{quaplectic field 2 equation}
\end{equation}

\noindent By using the commutation relations for ${\hat{Z}}_{a}^{\pm
}$ , it may be shown that
\[
\eta ^{b,c}\eta ^{a,d}{\hat{Z}}_{a,b}{\hat{Z}}_{c,d}=g_{2}( \hat{N})
=\hat{N}( \hat{N}-n) 
\]

\noindent This generalizes to $g_{k}( \hat{N}) ={\hat{N}( \hat{N}-n)
}^{k-1}.\ \ $Therefore, (86) may be written as 
\begin{equation}
\eta ^{b,c}\eta ^{a,d}{\hat{Z}}_{a}^{+}{\hat{Z}}_{b}^{-}{\hat{\varepsilon
}}_{c,d}\left. \left| \psi \right. \right\rangle  =f_{2}( c_{1},c_{2},d_{1},d_{2})
\left. \left| \psi \right. \right\rangle   
\end{equation}

\noindent where 
\begin{equation}
f_{2}( c_{1},c_{2},d_{1},d_{2}) =\frac{1}{2}\left( d_{2}-c_{2}-g_{2}(
d_{1}+c_{1}) \right) 
\end{equation}

\noindent This process may be repeated for higher order equations
yielding (70).\ \ The $c_{a}$ are labels for irreducible representations
and the $d_{a}$ are labels for states within these irreducible representations.

\appendix\label{lineelements}\label{O11}\label{canonical}\label{Ha}\label{Gb}\label{nullsurface}\label{atiunitary}\label{pct}\label{discrete}\label{
cover}\label{ordering}\label{anticommutator}\label{ht}\label{c01}\label{centralquaplectic}\label{casimirirrep}\label{lb}\label{gr}\label{spacetime}\label{blimit}\label{automorphisms}

\end{document}